# Dimensional crossover to a slow disordered spin bath below the mean spin spacing

Chikara Shinei[1,2,*]

[1] *Department of Applied Physics, Institute of Pure and Applied Sciences, University of Tsukuba, Tsukuba, Ibaraki 305-8573, Japan*
[2] *Japanese-French Laboratory for Semiconductor Physics and Technology J-FAST, CNRS, University Grenoble Alpes, Grenoble INP, University of Tsukuba, Tsukuba, Japan*
[*] shinei.chikara.fb@u.tsukuba.ac.jp

How thin must a layer of spins be to act as a two-dimensional bath? For nitrogen spins around an $NV^-$ centre in diamond, a cluster expansion with flip-flopping outer spins gives a crossover near the mean spin spacing. Thin layers follow two-dimensional scaling with a prefactor six to seven times that of bulk. Without hyperfine disorder the crossover remains but the step is under three: the disorder leaves each spin almost no resonant partner. Flip-flops allow several milliseconds, so another process must set the measured 0.5 ms; nitrogen spin–lattice relaxation is consistent.

Thin layers of spin defects are the working element of nanoscale magnetic resonance [1] and of programmable dipolar many-body systems [2,3]. In nitrogen δ-doped diamond, nitrogen-vacancy ($NV^-$) centres and the substitutional nitrogen (P1) spins that surround them are confined to a layer a few nanometres thick [4,5]. The P1 spins form the bath that limits $NV^-$ coherence. Hughes et al. measured a single-NV Hahn-echo time of 489–562 μs in a 3.55 nm $^{15}$N layer [5], seven times the value expected for bulk diamond of the same local nitrogen concentration [6]. Double electron–electron resonance in such layers shows the $\exp[-(t/T)^{2/3}]$ decay of a two-dimensional (2D) dipolar ensemble [2,5], and cluster-correlation-expansion (CCE) calculations reproduce a change in the concentration dependence of $T_2$ for thin layers [7].

This raises a simple question that has not been answered: at what thickness does a spin bath stop behaving as three-dimensional (3D)? The answer matters beyond geometry. A dipolar bath fluctuates through flip-flops of spin pairs whose energies match within their coupling $J \propto r^{-3}$. For $1/r^3$ couplings, three dimensions is the marginal case of the resonance-counting criterion for delocalization, and two dimensions is sub-critical [8–10]. A layer may therefore change not only the strength of the bath but its dynamics.

Here we determine the crossover thickness for the P1 bath, show that it lies close to the mean spin spacing, and identify the mechanism. Below the crossover the bath is slow enough that a process usually neglected, the spin–lattice relaxation of the P1 spins, can set the measured coherence.

*Model.*—The $NV^-$ centre sits in a uniform $^{15}$N layer of thickness w grown on (001), with its axis and a 322 G field along [111], as in Ref. [5]. Each P1 is an electron spin $\mathbf{S}_i$ and a nitrogen spin $\mathbf{I}_i$:

$$H = \sum_i H_i + \sum_{i<j} J_{ij}\left[S_i^z S_j^z - 1/4\left(S_i^+ S_j^- + \mathrm{h.c.}\right)\right], \tag{1}$$

$$H_i = \gamma_e B S_i^z + \boldsymbol{S}_i \cdot \mathsf{A}_i \cdot \boldsymbol{I}_i - \gamma_N B I_i^z,$$

with $J_{ij} = J_0(1 - 3\cos^2\theta_{ij})/r_{ij}^3$ and $J_0 = 52.1$ MHz nm$^3$; $H_i$ contains the electron and nuclear Zeeman terms and the hyperfine coupling. The hyperfine tensor $\mathsf{A}_i$ is oriented along one of four Jahn–Teller axes, which splits the bath into four spectral sub-groups. The $^{12}$C enrichment of Ref. [5] (99.998 %) leaves a $^{13}$C bath with an echo time of about 0.5 s, which is omitted [11].

*Method.*—We compute the Hahn echo $L(2\tau)$ by CCE at the pair level in units of whole P1 centres [7,12,13], with the dynamical-mean-field code dynacce of Ref. [14]. Each cluster is fully mixed. Every P1 outside it acts through the field of its electron, its mean field. Without this mean field, CCE underestimates the echo time of a bulk P1 bath two- to fourfold [15]. The question is how the mean field moves in time. We treat two cases. In the *frozen* limit, each outer P1 stays in one randomly drawn eigenstate, so its field is static. In the *pair flip-flop* (PFF) model, outer P1 centres closer than $4\ell$ flip-flop in pairs, conserving their total $S^z$; each flip-flop happens at a golden-rule rate set by the pair's present energy mismatch, including the fields of all other spins, and is sampled by kinetic Monte Carlo [11,14]. A pair whose energies do not match therefore stays frozen, and only resonant pairs move. The frozen limit leaves out these moves and so gives an upper bound on $T_2$. For bulk $^{14}$N P1 baths at 30 mT, PFF agrees with the measured relation $T_2 = 160$ μs·ppm/[N] [6] within 1.7 standard deviations, whereas the frozen limit overestimates it [14]. We therefore take PFF as the estimate and the frozen limit as the upper bound. Pairs farther apart than $4\ell$ do not contribute; enlarging the bath beyond the production radius shortens 2D echo times by about 10 %, less than their spread between configurations. Each layer is averaged over 8 configurations with 8 trajectories each [11].

Exact diagonalization checks the cluster expansion. For six P1 centres in a 2D sheet and in 3D, the frozen-limit CCE agrees with the exact echo within 0.025 up to 1000 coherence times [Figs. 1(b) and 1(c)], whereas CCE without the mean field does not. PFF decays faster than the exact echo in both geometries, because six spins cannot form a continuum into which a flip-flop can decay irreversibly. This test therefore validates the cluster expansion and the mean field, not the rates, which the bulk comparison above supports.

Two lengths describe a layer of local density $n_3$: the bulk spacing and the in-plane spacing,

$$\ell_3 = n_3^{-1/3}, \qquad \ell_2 = (n_3 w)^{-1/2}. \tag{2}$$

Since $w/\ell_2 = (w/\ell_3)^{3/2}$, the layer is thinner than $\ell_3$ exactly when it is thinner than $\ell_2$, so $w = \ell_3$ is the natural candidate for a crossover. For 2.14 ppm, the local concentration of Ref. [5], $\ell_3 = 13.8$ nm.

*Crossover thickness.*—Figure 1(a) shows the echo at 2.14 ppm for thicknesses from $8\ell_3$ to $0.07\ell_3$. For $w \geq 2\ell_3$ the decay is close to bulk. Below $\ell_3$ it moves to much longer times: the echo time grows from 0.27 ms at $\ell_3$ to 1.1 ms at $0.5\ell_3$ and 3.3 ms at the 3.55 nm of Ref. [5], against 59 μs in bulk.

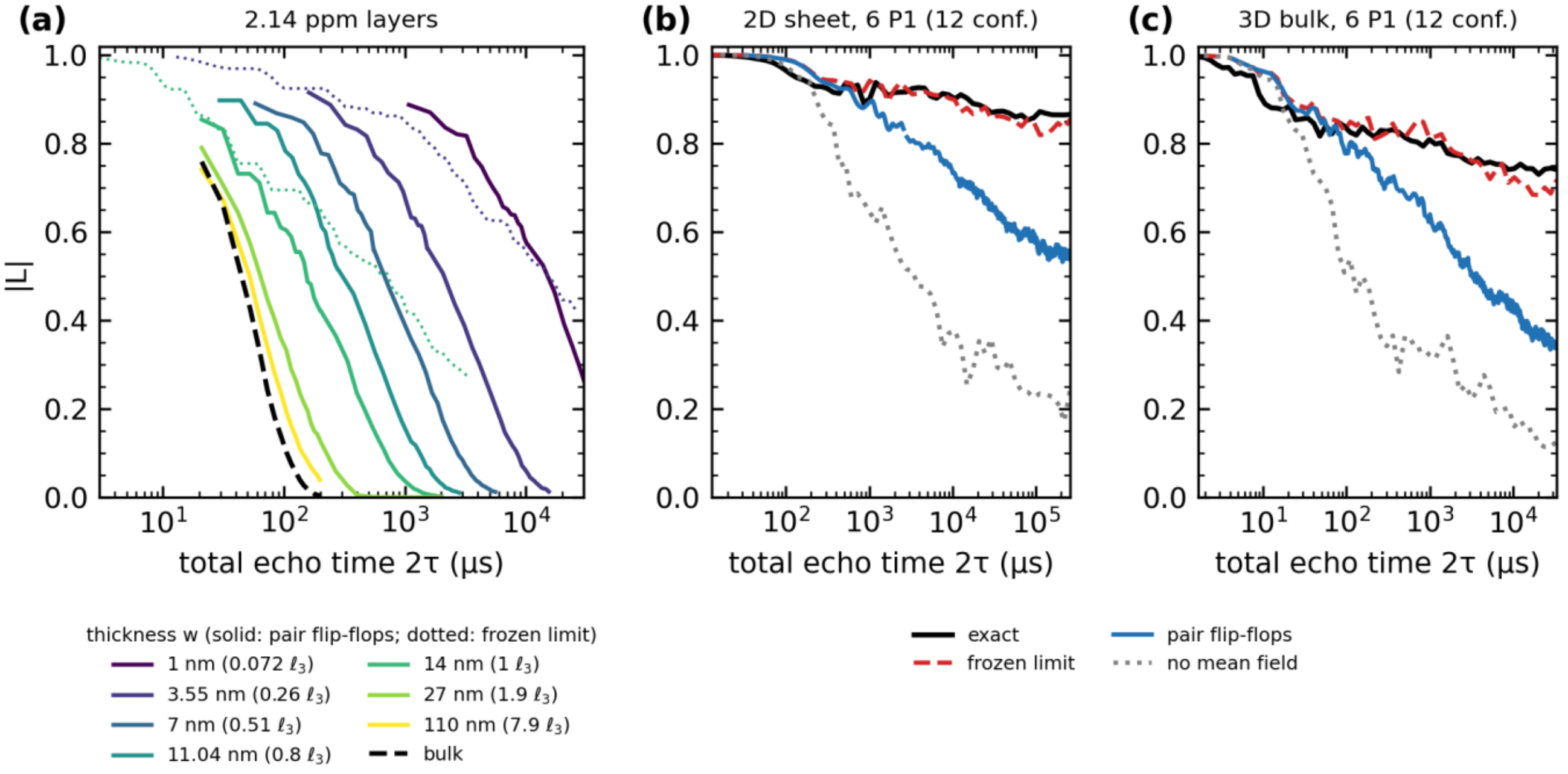

FIG. 1. (a) Hahn-echo coherence of an $NV^-$ centre in $^{15}N$ layers of local concentration 2.14 ppm ($\ell_3 = 13.8$ nm) and thickness w, and in bulk (dashed); PFF (solid) and frozen limit (dotted), 322 G. (b) Sheet and (c) bulk cluster of six P1 centres: exact diagonalization (black), frozen-limit CCE (red dashed), PFF (blue) and CCE without mean field (grey dotted), averaged over 12 configurations.

Figure 2(a) collects the echo time due to P1 flip-flops for local concentrations of 0.5, 2.14 and 10 ppm, spanning a factor of 20 in $\ell_3^3$. Relative to bulk, $T_2$ is 1.2–1.4 times longer at $w = 4$–$8\ell_3$ and 1.6–1.8 times at $2\ell_3$; this small gain is geometric, as fewer bath spins lie within reach of the $NV^-$. Below $\ell_3$ the gain is large and the same for all three concentrations: 18–19 times bulk at $0.5\ell_3$, 55–58 times at $0.25\ell_3$ and about 220 times at $0.1\ell_3$. In the frozen limit the echo of every layer thinner than $\ell_3$ stays above 1/e for the whole computed window of 100 coherence times (arrows), so there only the flip-flops bring the echo to 1/e.

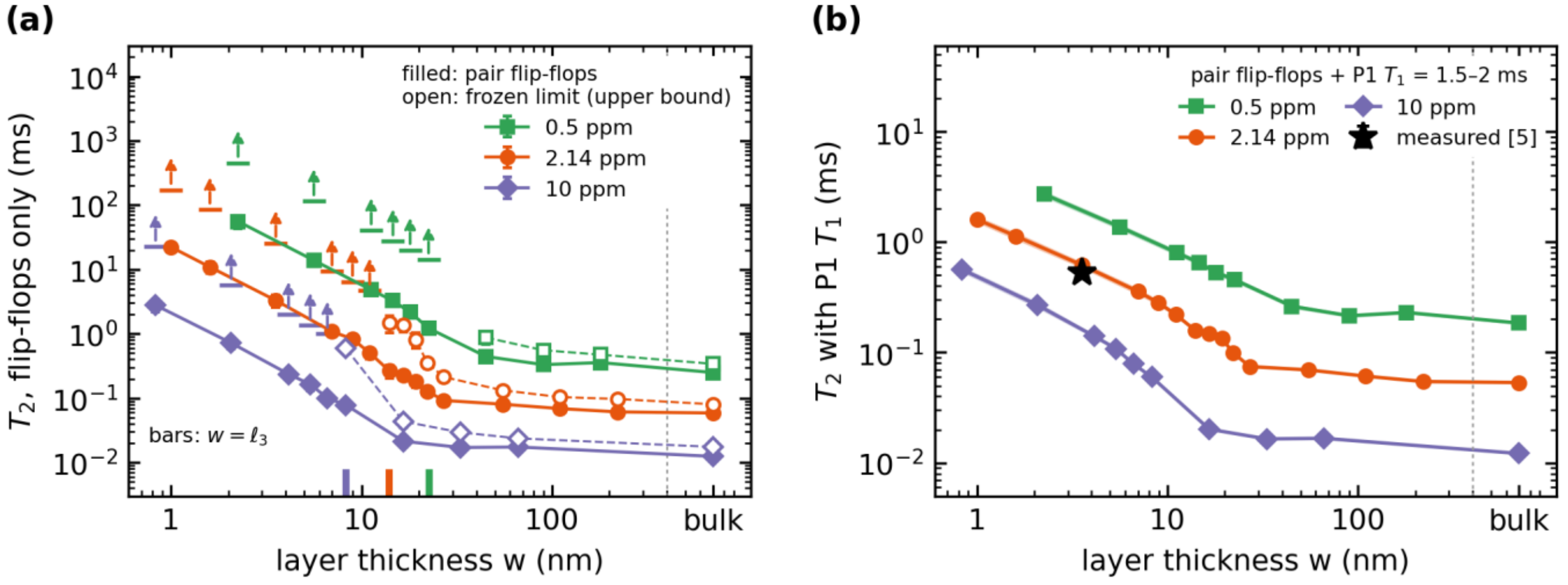

FIG. 2. Echo time ($|L| = 1/e$) versus layer thickness for local concentrations of 0.5, 2.14 and 10 ppm; bulk at the right. Bars at the bottom mark $w = \ell_3$. (a) P1 flip-flops only: PFF (filled) and frozen limit (open). Arrows: lower bounds where the frozen-limit echo stays above 1/e to the end of the computed window (100 coherence times). Error bars: bootstrap over configurations. (b) PFF including spin–lattice relaxation of the P1 spins with $T_1$ = 1.5–2 ms (bands) [16]. Star: measured $T_2$ of Ref. [5].

Thin layers follow a simple law. A dipolar bath has one energy scale, $J_0/\ell^3$ with $\ell$ the mean spacing, so $T_2 \propto \ell^3$, or $T_2 = \kappa/c_\ell$ with $c_\ell$ the concentration of a bulk crystal of spacing $\ell$. In bulk, $\ell = \ell_3$ and $\kappa_3 = 0.125$ ms·ppm for all three concentrations. For $w \lesssim 0.65\ell_3$ the spacing that matters is $\ell_2$, and the 23 thin layers computed at up to 10 ppm, with areal densities from 1.1 to 54 ppm·nm, give $\kappa_2 = 0.88 \pm 0.03$ ms·ppm [11]. In terms of the areal density $\sigma$,

$$T_2^{2\mathrm{D}} \simeq 3.1\ \mathrm{ms} \left(\frac{\sigma}{7.6\ \mathrm{ppm} \cdot \mathrm{nm}}\right)^{-3/2}, \tag{3}$$

independent of thickness. The crossover between the two laws is at

$$w^* \simeq \ell_3 = n_3^{-1/3}, \tag{4}$$

the thickness below which a layer holds, on average, less than one spin per in-plane spacing along its normal. This is the criterion Hughes et al. used to define a 2D layer [5]. The change is gradual; at 2.14 ppm its midpoint on a logarithmic scale is at $1.47 \pm 0.11\ell_3$ [11]. The step from $\kappa_3$ to $\kappa_2$ is a factor of six to seven once the bath-size correction of Ref. [11] is allowed for. Not all of it is dynamics. Without the mean field, where nothing detunes a pair, the step is already $2.9 \pm 0.4$ at 2.14 ppm, which we attribute to the different geometry of two and three dimensions. The remaining factor, $2.5 \pm 0.4$ at this concentration, is the change of dynamics: the 2D bath is slower than its spacing alone implies.

The long-time behaviour shows the same change. Figure 3(a) shows the coherence left after 50 coherence times. In the frozen limit it is 0.3–0.6 for every layer with $w \leq \ell_2$, including the four layers of Ref. [5], 0.1–0.2 at $2\ell_2$ and below 0.04 beyond $4\ell_2$: below $\ell_2$ the static fields alone cannot dephase the $NV^-$. With flip-flops the coherence has decayed by then ($|L| \leq 0.04$). The thin-layer bath is therefore slow, not frozen.

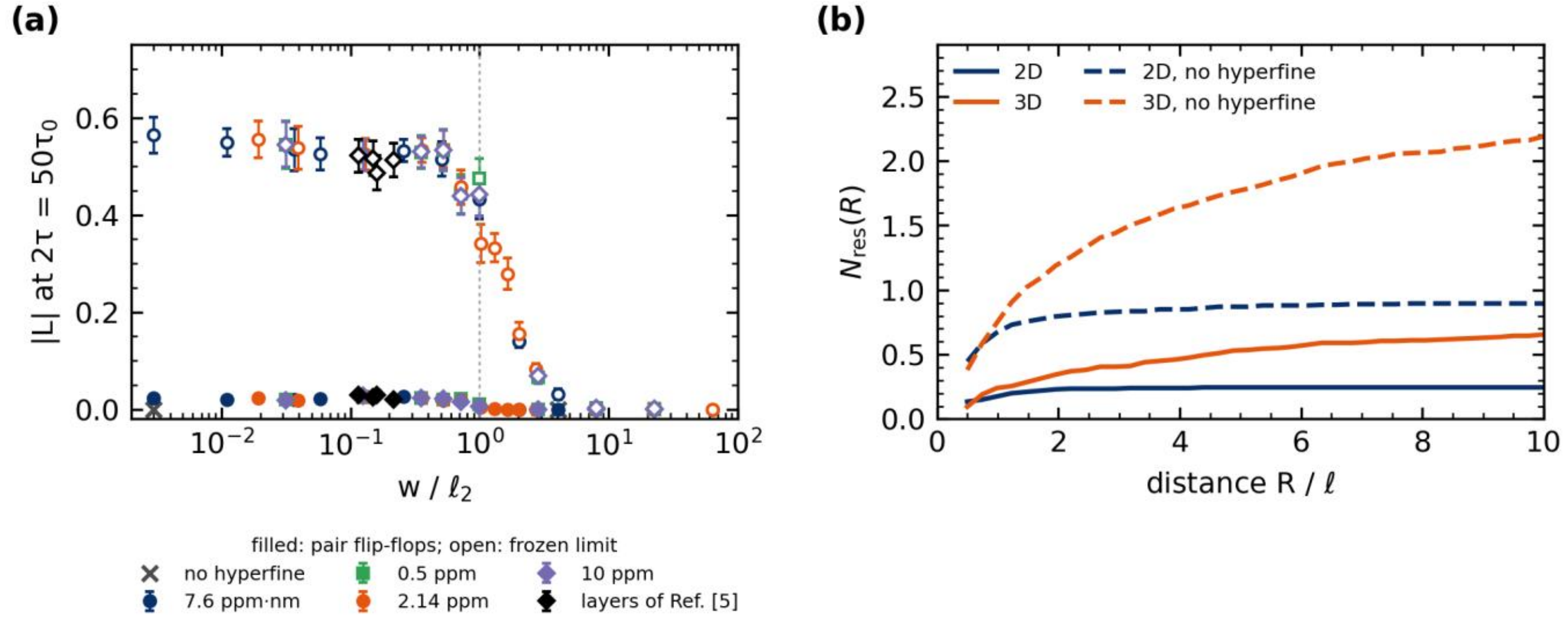

FIG. 3. (a) Coherence left after 50 coherence times versus $w/\ell_2$ in the frozen limit (open) and with PFF (filled), for a fixed dose of 7.6 ppm·nm, fixed local concentrations of 0.5, 2.14 and 10 ppm, the four layers of Ref. [5] (diamonds), and bare electrons without hyperfine (crosses). A sheet is plotted at $w/\ell_2 = 3 \times 10^{-3}$. (b) Mean number of resonant partners [Eq. (5)] of a P1 within distance R (units of the mean spacing $\ell$) in a 2D sheet and in 3D bulk, with and without hyperfine sub-groups.

*Mechanism.*—A flip-flop of spins i and j requires them to belong to the same hyperfine sub-group and

$$\frac{|J_{ij}|}{2} > |h_i - h_j|, \qquad h_i = \sum_{k \neq i,j} J_{ik}\, S_k^z, \tag{5}$$

where $h_i$ is the static field of the rest of the bath. With a width W of the distribution of $h_i - h_j$, the number of resonant partners within distance R scales as

$$N_{\mathrm{res}}(R) \sim \int^R n_d \frac{J_0}{W r^3}\, d^d r \ \propto \begin{cases} \ln R, & d = 3, \\ \mathrm{const.}, & d = 2, \end{cases} \tag{6}$$

the resonance-counting argument of Refs. [8–10]. Figure 3(b) counts the pairs that satisfy Eq. (5) for random P1 configurations. In 3D the number grows from 0.38 at $2\ell$ to 0.66 at $10\ell$; in 2D it saturates at 0.25 by $3\ell$. In a thin layer, then, a P1 has on average a quarter of a resonant partner at any distance, so resonant pairs are isolated rather than connected, and the flip-flops that dephase the $NV^-$ are rare. This accounts for the part of $\kappa_2$ beyond geometry. Removing the hyperfine sub-groups triples the counts and shortens the echo time of a sheet sixfold, from 3.2 to 0.52 ms (PFF) [11]. A thickness series of such bare electrons at 2.14 ppm separates the two ingredients [11]. The midpoint of the step stays at $1.5\ell_3$ (16–84 % range $1.4$–$1.6\ell_3$), so the crossover does not need the hyperfine disorder. The step falls to $2.7 \pm 0.8$, only $1.3 \pm 0.3$ times that without the mean field, so no slowing is resolved. The thickness at which a bath becomes two-dimensional thus follows from its spacing, and its slowness from the sub-groups. The crossover at $w \approx \ell_3$ is where the bath spins within a resonance-relevant distance stop filling three dimensions.

*Measured coherence.*—Below w*, flip-flops alone permit echo times of several milliseconds at the density of Ref. [5] [Eq. (3)], and beyond 10 ms for $w \le 0.25\ell_3$ at 0.5 ppm [Fig. 2(a)], six times or more beyond the measured 0.5 ms. Another process must therefore terminate the echo. A natural candidate is spin–lattice relaxation of the P1 spins, with $T_1 \approx 1.5$–2 ms at room temperature in synthetic diamond [16]. A random flip of a P1 electron shifts the field at the $NV^-$ and is not refocused. We add it as an independent two-state process for each P1, evaluated exactly and multiplied by the CCE echo configuration by configuration [11]:

$$L(2\tau) = L_{\mathrm{CCE}}(2\tau) \prod_j \langle e^{i\phi_j} \rangle, \tag{7}$$

$$\phi_j = a_j \int_0^{2\tau} s(t)\, S_j^z(t)\, dt,$$

where $a_j$ is the $NV^-$–P1 coupling and $s(t) = \pm 1$ changes sign at $\tau$. In bulk this lowers $T_2$ by at most 28 % (10 % at 2.14 ppm). In 2D layers the effect is large [Fig. 2(b)]: $T_2$ falls to 0.13–2.9 ms for $w \le 0.5\ell_3$, and for the 3.55 nm layer of Ref. [5] to 0.58–0.67 ms, against 489–562 μs measured. NV ensembles in irradiated spots of the same sample have $T_2 \approx 0.1$ ms, which Ref. [5] attributes mainly to the P1 and $NV^-$ spins, with a contribution from other defects. Interactions among NVs of the probe orientation are not refocused by the echo; we therefore compare with single NVs in unirradiated regions. The measured single-NV echo lies within the predicted range up to about 0.7 ms and decays somewhat faster beyond [Fig. 4(a)]. Because the P1 $T_1$ was measured in bulk crystals and not in the layer, we regard this agreement as indicating, not proving, that bath relaxation limits present δ-doped layers. A direct test is temperature: since $1/T_1 \propto T^5$ above 80 K [16], $T_2$ of a 2D layer should rise on cooling, to 1.0–1.8 ms at 200 K [Fig. 4(b)], until it approaches the flip-flop limit of Eq. (3), whereas a flip-flop-limited echo would not depend on temperature.

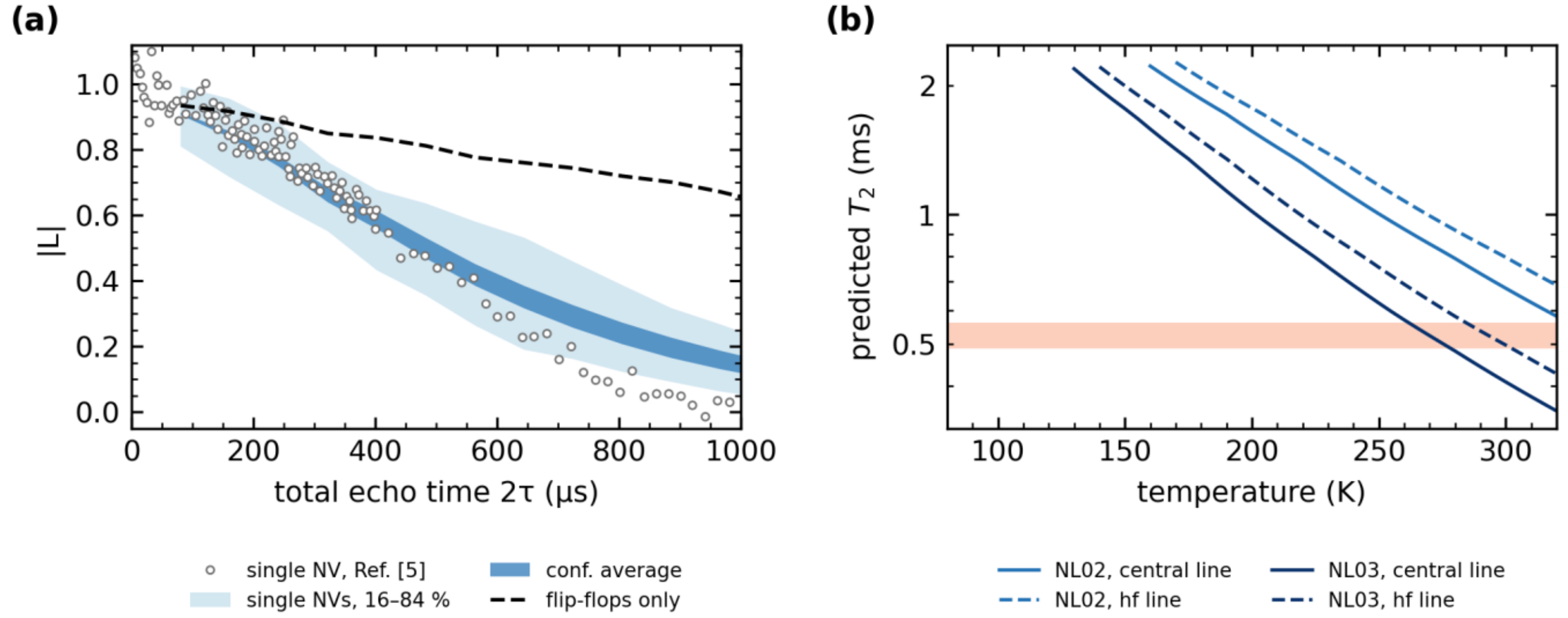

FIG. 4. The 3.55 nm, 7.6 ppm·nm layer of Ref. [5]. (a) Single-NV echo of Ref. [5] (circles; data points read from the plot in its Fig. 3(c) and replotted), compared with Eq. (7) for $T_1 = 1.5$–2 ms: configuration average (dark band) and 16–84 % range of single-NV curves (light band). Dashed: P1 flip-flops only (PFF). (b) Echo time predicted from the temperature-dependent P1 $T_1$ of Ref. [16] (synthetic crystals NL02, 25 ppm, and NL03, 95 ppm; central and hyperfine lines). Band: measured $T_2$ at room temperature.

*Discussion.*—Equations (3) and (4) give a simple design rule. It quantifies, as a threshold for $T_2$, the criterion used in Ref. [5] to identify a 2D layer: a thickness smaller than the P1 spacing (13.8 nm there, which is $\ell_3$ at 2.14 ppm). For $w \ge 2\ell_3$, thinning a δ-doped layer at fixed local concentration lengthens $T_2$ by at most twofold. The range between $2\ell_3$ and $0.65\ell_3$ is the transition, with $\ell_3$ inside it: there $T_2$ is already 4.5–6.3 times its bulk value. For $w \lesssim 0.65\ell_3$, $T_2$ is set by the areal density alone, with a prefactor six to seven times that of bulk. The flip-flops are then so slow that, at the concentrations of Fig. 2, usually neglected processes such as bath spin–lattice relaxation shorten $T_2$ at room temperature. Cooling, which lengthens $T_1$, should then raise $T_2$ towards the several milliseconds that the 2D bath permits [Fig. 4(b)], whereas in bulk, where temperature-independent flip-flops set $T_2$, it should barely change. The temperature dependence of $T_2$ is therefore a direct test of a 2D bath. The crossover is the same at 100 ppm [11], but in dense layers $NV^-$ centres and other nitrogen defects may limit $T_2$ first.

The P1–NV system meets four conditions that a slow 2D bath and its detection require. The layer can be grown thinner than the mean spacing, as in nitrogen δ-doping. The splitting between spectral sub-groups, at least 20 MHz, exceeds the dipolar coupling $J_0/\ell^3$ of 3–20 kHz at 2.14 ppm by three to four orders of magnitude. The probe is far off resonance with the bath at 322 G, so it senses dephasing without exchanging spin with it. Isotopic purification removes the competing nuclear bath. At room temperature P1 spin–lattice relaxation still limits the echo; cooling should lift it. These conditions, rather than diamond itself, indicate where else the effect should be sought.

The crossover thickness survived the removal of the hyperfine disorder and should carry over to other dipolar baths. The slowness requires a static disorder that splits the spins into groups out of resonance with each other, as hyperfine coupling and the Jahn–Teller axes do for P1. It should appear in baths with such a splitting, for example dense $NV^-$ layers with several axes and spin defects with strong hyperfine coupling in van der Waals materials [17]. For surface spins without such a splitting [3] we expect the crossover but not the slowing. Our calculation is limited to pair clusters, checked by exact diagonalization of small systems; higher orders diverge for electron baths, as in Ref. [7].

*Acknowledgments.*—This study was funded by JSPS KAKENHI Grant Number 25K17636.

# Supplemental Material: Dimensional crossover to a slow disordered spin bath below the mean spin spacing

Chikara Shinei

*Reference and equation numbers are those of the main text.*

## S1. Bath model

**Geometry.** Supplementary Fig. S1 sketches the system. The layer is grown on (001); the $NV^-$ axis and the 322 G field are along [111], at 54.7° to the layer normal. P1 centres are placed uniformly in a slab of thickness w with areal density σ (ppm·nm) or local concentration c = σ/w, the $NV^-$ depth being uniform within the slab; a sheet (w = 0) and bulk diamond are the limits. For the layers of Ref. [5], w is the full width at half maximum of the $^{15}N$ depth profile measured by secondary-ion mass spectrometry, 3.55(0.04) nm for sample A, which Ref. [5] notes is convolved with the surface roughness ($S_q$ < 0.5 nm) and may be an upper bound; the areal density σ = 7.6 ± 1.8 ppm·nm is the P1 density from NV–DEER, about eight times smaller than the total nitrogen from SIMS. Ref. [5] defines two-dimensionality as a thickness smaller than the inter-spin spacing, which for sample A it gives as ≈13.8 nm; this is $\ell_3$ at the local concentration σ/w = 2.14 ppm, so their criterion is w < $\ell_3$, the crossover found here. The characteristic spacing is $\ell = \ell_2$ for $w \le \ell_2$ and $\ell = (w/n_2)^{1/3}$ otherwise ($n_2$ the areal density); for bulk $\ell = \ell_3$. Spins are kept within $r_{bath} = \max(5, r_{dip}/\ell + 1.5)\,\ell$ of the $NV^-$ and outside 0.5 nm.

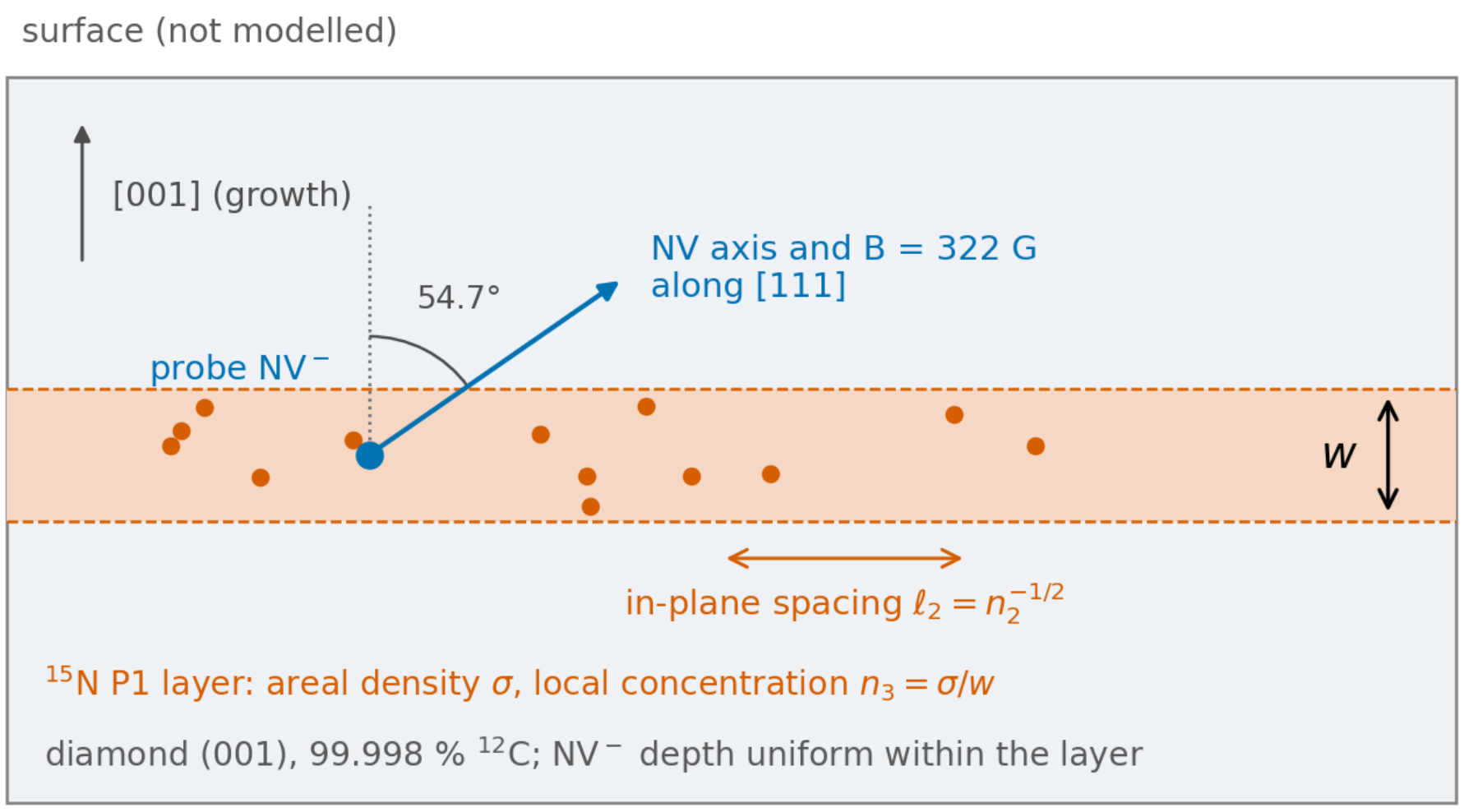

FIG. S1. Geometry of the model (not to scale). The $^{15}N$ P1 centres (orange) are distributed uniformly in a layer of thickness w grown on (001); the probe $NV^-$ (blue) lies at a uniformly distributed depth within the layer. The $NV^-$ axis and the 322 G field are along [111], at 54.7° to the layer normal. The bath is characterized by the areal density σ, the local concentration $n_3 = \sigma/w$, the in-plane spacing $\ell_2 = n_2^{-1/2}$ and the bulk spacing $\ell_3 = n_3^{-1/3}$.

**Spins.** Each P1 is an electron (S = 1/2) and a $^{15}N$ nucleus (I = 1/2) coupled by $\mathsf{A} = A_\perp \mathbb{1} + (A_\parallel - A_\perp)\hat{n}\hat{n}^T$ along a Jahn–Teller axis $\hat{n}$ drawn from the four ⟨111⟩ directions; the principal values are those of $^{14}N$ (114.03, 81.31 MHz) scaled by $\gamma(^{15}N)/\gamma(^{14}N) = -1.4028$. All spins interact by point dipoles. The $NV^-$ (D = 2.87 GHz) qubit is $|m_s = 0\rangle \leftrightarrow |m_s = -1\rangle$.

**$^{13}C$ bath.** The layers of Ref. [5] are grown with 99.998 % $^{12}C$, leaving 20 ppm of $^{13}C$. We computed the Hahn echo of the $NV^-$ in a random 20 ppm $^{13}C$ bath alone, by conventional CCE2 with a thermal nuclear bath (dynacce), averaging four configurations. The echo time (1/e) is 511 and 516 ms for pair cutoffs of 200 and 300 Å (bath radii 500 and 600 Å), with |L| > 0.998 up to 30 ms. The $^{13}C$ bath is therefore omitted.

## S2. Cluster expansion and the two mean fields

All calculations use dynacce [14]. The Hahn echo is computed by CCE in units of whole P1 centres, with clusters of one centre and of two centres closer than $r_{dip}$ [7]. Inside each cluster the bath is fully mixed, which averages exactly over its internal states. Every P1 outside a cluster enters through the electron $\langle S_z \rangle$ of its present electron–nuclear eigenstate, the mean field. The initial eigenstates are drawn with equal probability; each draw is one trajectory.

**Frozen limit.** The outer eigenstates keep their initial values for the whole echo, as in the mean-field average of Refs. [7,15]. This echo is computed for all pairs without screening, on 61 logarithmic points from $0.05\tau_0$ to $100\tau_0$.

**Pair flip-flops (PFF).** Outer P1 centres closer than $4\ell$ to one another form flip-flop pairs. An event is a joint transition (a, b) → (a′, b′) of the eigenstates of a pair (k, j) that changes both, at the golden-rule rate

$$W = 2\pi\,|V|^2\,g(\Delta;\sigma^2), \qquad \sigma^2 = \sigma_0^2 + (2|V|)^2, \qquad (S1)$$

with V the flip-flop matrix element between the two eigenstate pairs and g a normalized Gaussian of the energy mismatch

$$\Delta = (E_{a'} - E_a) + (E_{b'} - E_b) + \delta\boldsymbol{x}_k \cdot \boldsymbol{h}_k + \delta\boldsymbol{x}_j \cdot \boldsymbol{h}_j + \delta\boldsymbol{x}_k \cdot \mathsf{G}_{kj} \cdot \delta\boldsymbol{x}_j, \qquad (S2)$$

where $h_k$ is the present mean field of all other P1 centres at k and δx the change of the electron ⟨**S**⟩. The term 2|V| caps the rate of strongly coupled pairs at about |V|; $\sigma_0 = 0$. Transitions whose intrinsic energy change exceeds 10 MHz are excluded, which removes non-secular double flips of close pairs. Because Δ is the actual mismatch, a detuned pair does not flip, and the frozen limit is recovered when the rates vanish. Events are drawn by kinetic Monte Carlo with thinning: candidates are drawn from the Gaussian-free bound of W, which depends only on the pair's own states, and accepted with probability $\exp(-\Delta^2/2\sigma^2)$. After each event the mean fields are updated. The $NV^-$ coherence of each cluster is then computed in the resulting time-dependent field, and the cluster contributions are multiplied as in CCE.

**Screening and time windows.** Pairs whose echo is insensitive to the mean field are kept static; the remaining pairs (about $1.4 \times 10^3$ in the short window and $1.2 \times 10^4$ in the long window of a thin layer) are computed with the time-dependent field. The screening is based on the sensitivity up to 60 % of each window, so only the first 60 % of each window is used. The PFF echo is computed on a uniform grid in two windows, 0–$10\tau_0$ (32 points) and 0–$100\tau_0$ (50 points); the long window is computed only when $|L(10\tau_0)| > 0.05$.

**Statistics and time scale.** Each layer is averaged over 8 positional configurations (seeds shared between layers and between the two mean fields) with 8 trajectories each; the echo is the modulus of the averaged complex coherence, with a running minimum applied before characteristic times are read. Errors are bootstrap standard deviations over configurations. Coherence times are compared with $\tau_0 = (71.6\ \mu s\,\mathrm{ppm})/c_\ell$, where $c_\ell = \ell^{-3}/n_C$ is the concentration of a bulk crystal of mean spacing ℓ ($n_C = 175.9\ \mathrm{nm}^{-3}$); for σ = 7.6 ppm·nm, $\ell_2 = 27.3$ nm and $\tau_0 = 257$ µs. In bulk $^{15}N$ diamond the echo time is $T_2 \cdot c = 124$–126 µs·ppm with PFF and 170–175 µs·ppm in the frozen limit.

## S3. Convergence in the pair cutoff

The production cutoffs are $r_{dip} = 8\ell$ for $w \le \ell_2$, $6\ell$ for $\ell_2 < w \le 16\ell_2$ and $4\ell$ beyond, and the bath extends to $R_{bath} = \max(5\ell, r_{dip} + 1.5\ell)$ from the $NV^-$. Supplementary Fig. S2 tests both lengths with the PFF echo, averaged over four configurations.

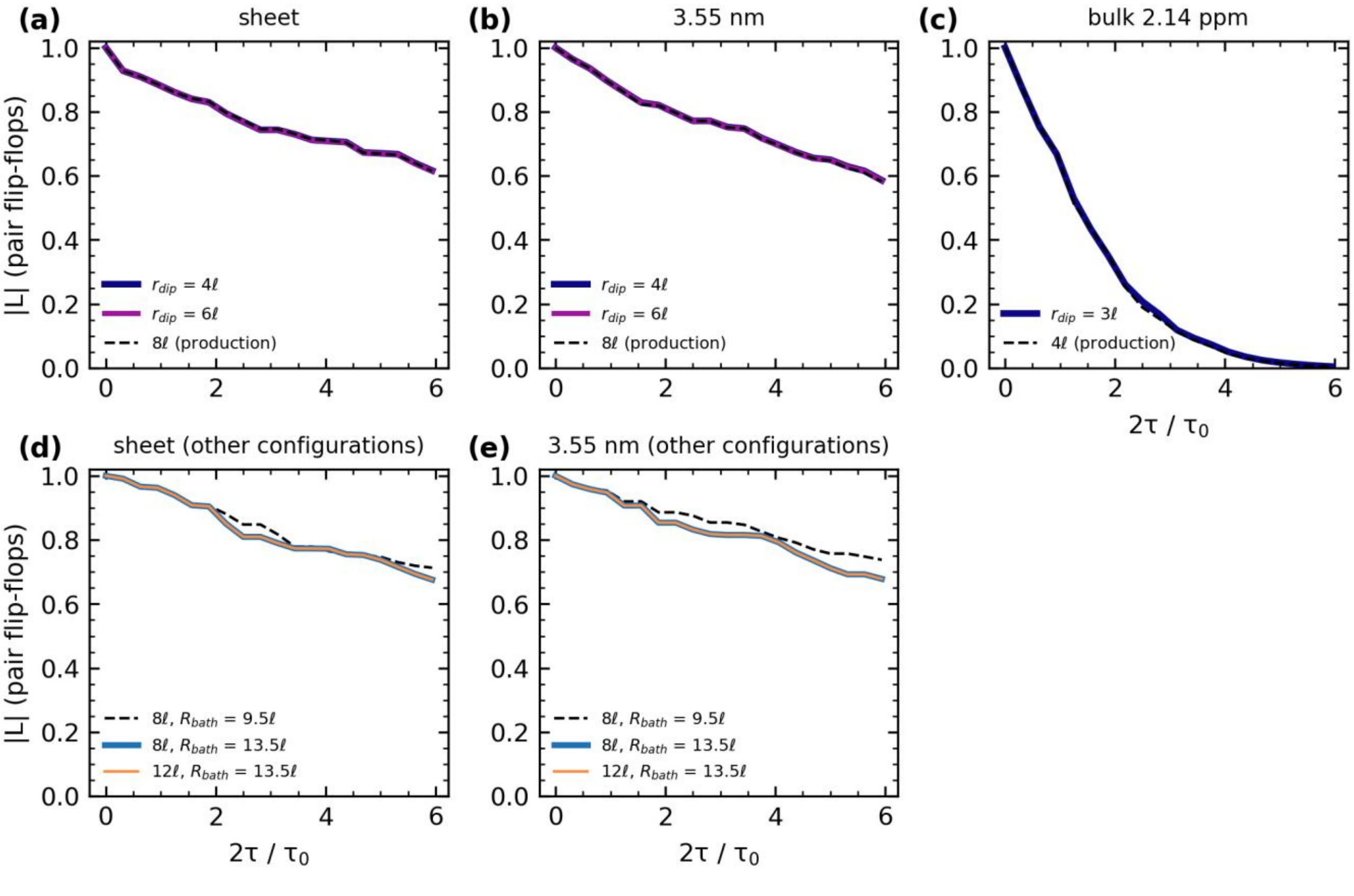

FIG. S2. Convergence of the PFF echo. (a)–(c) Pair cutoff $r_{dip}$ (colours) for the sheet and the 3.55 nm layer (σ = 7.6 ppm·nm) and bulk at 2.14 ppm, with the same configurations; dashed black: production cutoff. (d), (e) Bath radius for four other configurations generated within 13.5ℓ: $r_{dip} = 8\ell$ with $R_{bath} = 9.5\ell$ (production, dashed) and 13.5ℓ (blue), and $r_{dip} = 12\ell$ with $R_{bath} = 13.5\ell$ (orange).

**Pair cutoff.** With the same positions, Jahn–Teller axes and initial states, the echo for $r_{dip} = 4$ and $6\ell$ coincides with the production $8\ell$ for the sheet and the 3.55 nm layer at $\sigma = 7.6$ ppm·nm [Figs. S2(a) and S2(b)], and $r_{dip} = 3\ell$ coincides with the production $4\ell$ in bulk at 2.14 ppm [Fig. S2(c)]. Pairs farther apart than about $4\ell$ therefore do not contribute.

**Bath radius.** Enlarging $R_{bath}$ changes the random positions near the $NV^-$ as well, so we generated four further configurations within $13.5\ell$ and cut the same bath at the production radius of $9.5\ell$ [Figs. S2(d) and S2(e)]. In the $13.5\ell$ bath, $r_{dip} = 8$ and $12\ell$ again give the same echo. Cutting the bath at $9.5\ell$ slows the decay: the times at which $|L|$ falls to 0.8, 0.75 and 0.7 are longer by 0–20 % (about 10 % on average) than in the $13.5\ell$ bath. Spins beyond $9.5\ell$ thus add a small amount of dynamics to the bath near the $NV^-$. The 2D echo times and $\kappa_2$ may accordingly be overestimated by about 10 %, which is smaller than their spread between configurations (±17 % for eight configurations). A reduction of $\kappa_2$ by 10–20 % would leave it 5.5–6.5 times $\kappa_3$; it would not move the crossover at $w \approx \ell_3$, and it would change the $T_1$-limited echo times of main-text Fig. 2(b) only weakly, since those are set mainly by $T_1$.

## S4. Exact diagonalization of small clusters

Six P1 centres are placed in a $\sigma = 7.6$ ppm·nm sheet (2D) or in bulk at 2.14 ppm (3D), the six nearest to the $NV^-$ being kept. The bath Hamiltonian is Eq. (1) of the main text with the full hyperfine tensors. The $NV^-$ adds $-\sum_j a_j S_j^z$ for $m_s = -1$, $a_j$ the $NV^-$–electron dipolar coupling. The two Hamiltonians (dimension 4096) are diagonalized exactly, and $L(2\tau) = \mathrm{Tr}[(U_0 U_1)^\dagger U_1 U_0]/D$ is evaluated with eight random vectors. The same centres and Jahn–Teller axes are then treated by dynacce CCE2 with all pairs: without mean field, in the frozen limit and with PFF. Averaged over 12 configurations:

| | $2\tau/\tau_0$ | exact | frozen limit | PFF | no mean field |
|---|---|---|---|---|---|
| 2D | 10 | 0.917 | 0.924 | 0.822 | 0.516 |
| 2D | 1000 | 0.866 | 0.855 | 0.548 | 0.243 |
| 3D | 10 | 0.831 | 0.845 | 0.725 | 0.365 |
| 3D | 1000 | 0.741 | 0.718 | 0.348 | 0.126 |

The frozen-limit CCE agrees with the exact echo within 0.025 in both geometries, so the pair expansion with a mean field is accurate. PFF decays faster: six spins have a discrete spectrum, so a flip-flop oscillates back and forth instead of decaying into a continuum, while PFF treats every flip-flop as incoherent. This test therefore does not probe the PFF rates, which are supported by the bulk comparison of Ref. [14].

**CCE3.** Adding all P1 triples within $2\ell$ of one another to the $8\ell$ pair expansion of the sheet (frozen limit) produces $|L| > 1$ (1.08 and 1.16 at $3\tau_0$ and $10\tau_0$, with single samples up to 18), the known divergence of higher orders for electron baths [7].

## S5. Echo time versus thickness and the 2D scaling

Supplementary Table S1 lists the echo times plotted in main-text Fig. 2 and those of the 100 ppm series described below. For each local concentration, the 1/e time is given where it is reached within the computed window, and otherwise the window end as a lower bound.

Supplementary Fig. S3 plots $T_2 c_\ell$, the echo time multiplied by the concentration of a bulk crystal with the layer's mean spacing $\ell$, against $w/\ell_3$ for all computed layers. With PFF the values collapse onto two plateaus: $\kappa_3 = 0.124$–$0.126$ ms·ppm in bulk and $\kappa_2 = 0.88 \pm 0.03$ ms·ppm (mean and standard deviation of 23 layers with $w \leq 0.65\ell_3$, $\sigma = 1.1$–$54$ ppm·nm), with the change between about $0.65\ell_3$ and $2\ell_3$. Since $c_\ell = \ell_2^{-3}/n_C = n_C^{1/2}(\sigma \cdot 10^{-6})^{3/2}$ for $w \leq \ell_2$, $\kappa_2$ gives Eq. (3) of the main text. Supplementary Fig. S4 shows the time at which $|L|$ reaches 0.6, relative to bulk of the same local concentration, against $w/\ell_3$: it is defined for all layers in both mean fields, and the three concentrations collapse onto one curve.

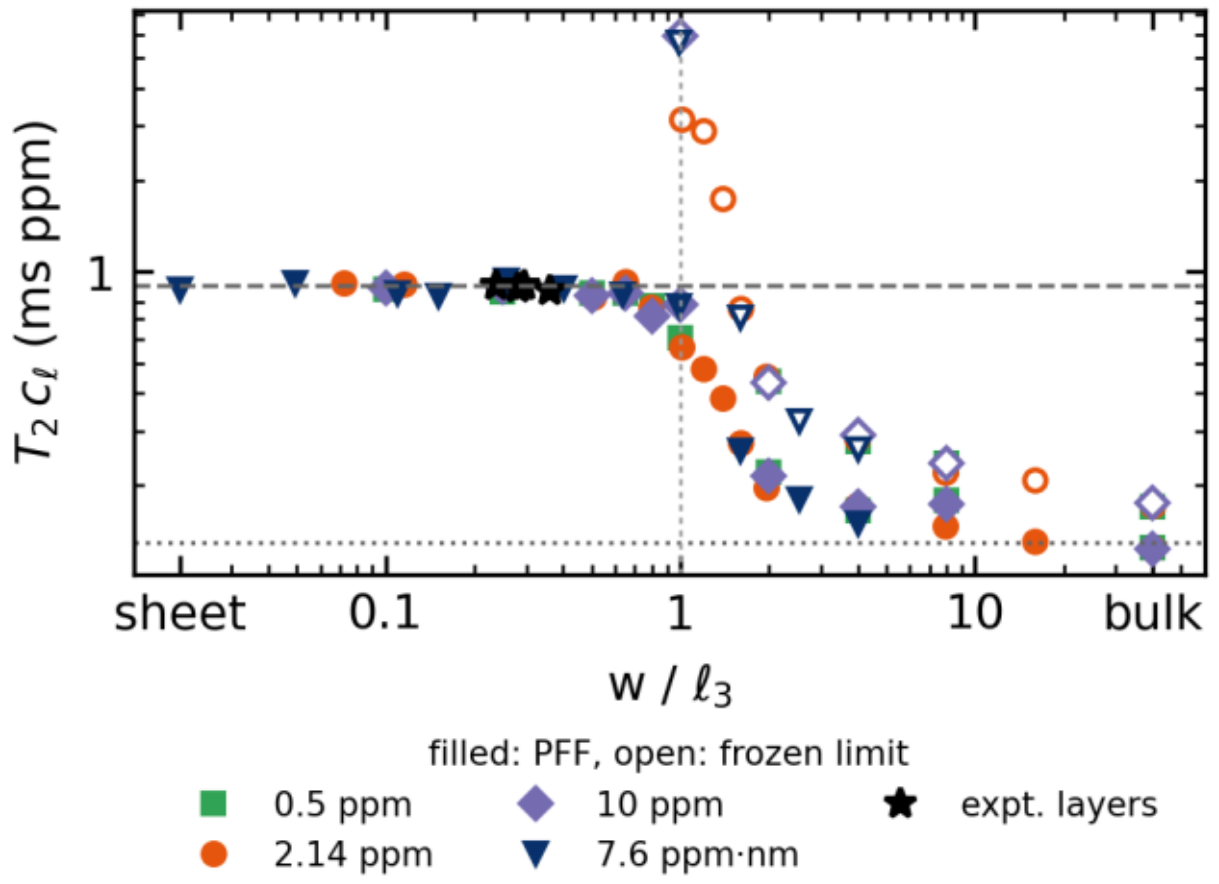

FIG. S3. Echo time multiplied by the concentration $c_\ell$ of a bulk crystal with the layer's mean spacing, versus $w/\ell_3$, for fixed local concentrations of 0.5, 2.14 and 10 ppm, a fixed dose of 7.6 ppm·nm and the four layers of Ref. [5]; PFF (filled) and frozen limit (open, where $|L| = 1/e$ is reached in the window). Dashed: $\kappa_2 = 0.9$ ms·ppm; dotted: $\kappa_3 = 0.13$ ms·ppm.

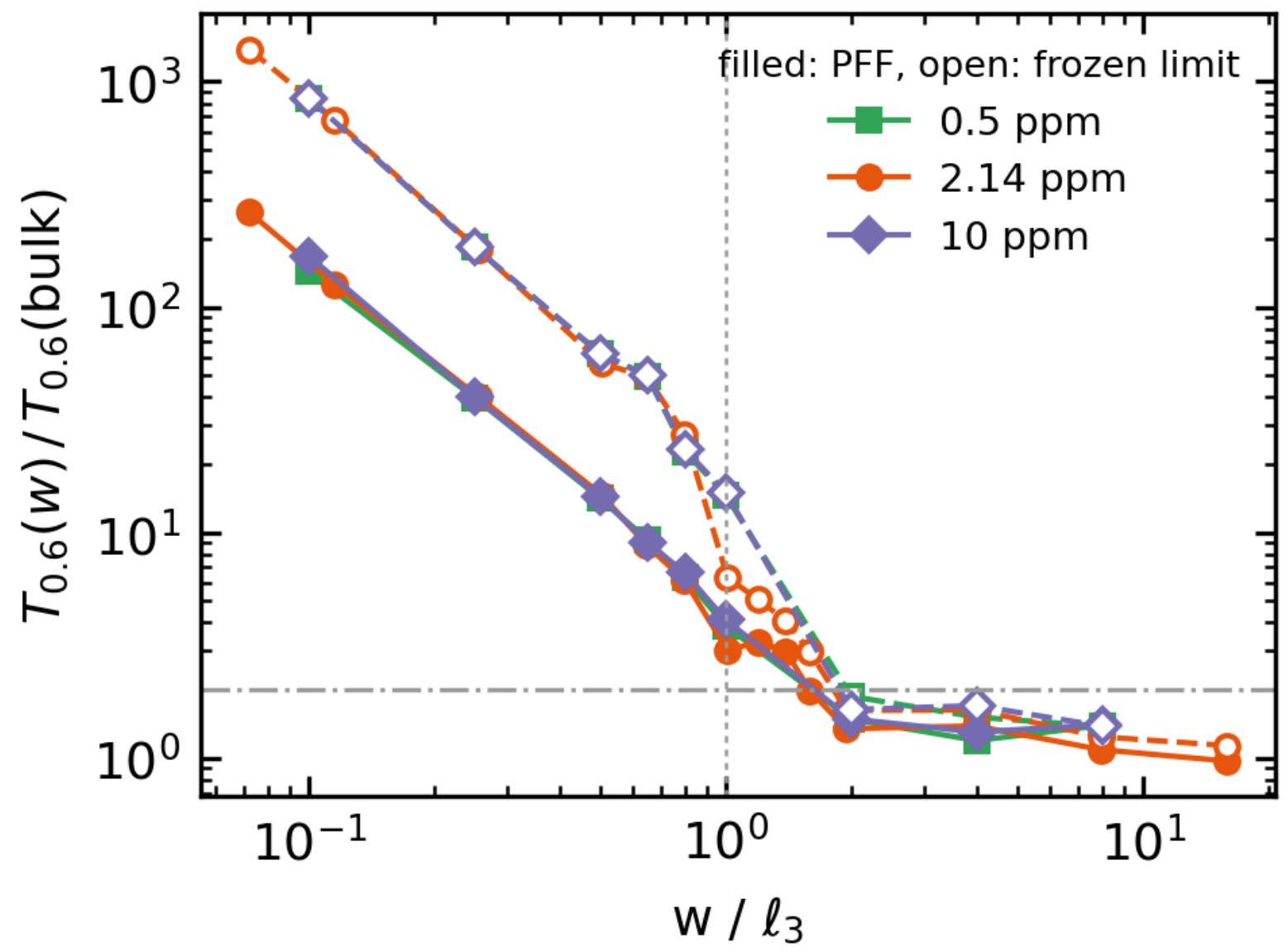

FIG. S4. Time at which $|L|$ reaches 0.6, relative to bulk of the same local concentration, versus $w/\ell_3$; PFF (filled) and frozen limit (open). Dash-dotted: factor 2; dotted: $w = \ell_3$.

Supplementary Fig. S5 repeats the thickness series at 100 ppm ($\ell_3$ = 3.84 nm; $w = 0.1$–$8\ell_3$ and bulk, eight configurations each; not included in $\kappa_2$ above). The ratio $T_2(w)/T_2(\mathrm{bulk})$ follows the 10 ppm series within about 10% at every $w/\ell_3$ and is 6.3 at $w = \ell_3$ for both; in the frozen limit the ratio at $w = \ell_3$ is 34 for both. The bulk echo time, 1.21 µs, gives $T_2c$ = 0.121 ms·ppm, close to $\kappa_3$. Within the P1 model the crossover is therefore set by $w/\ell_3$ alone, whatever the concentration. At such concentrations, however, the echo of a real layer may be limited first by $NV^-$ centres and other nitrogen-related defects, which the model omits. Supplementary Fig. S6 shows main-text Fig. 2 with the 100 ppm series added.

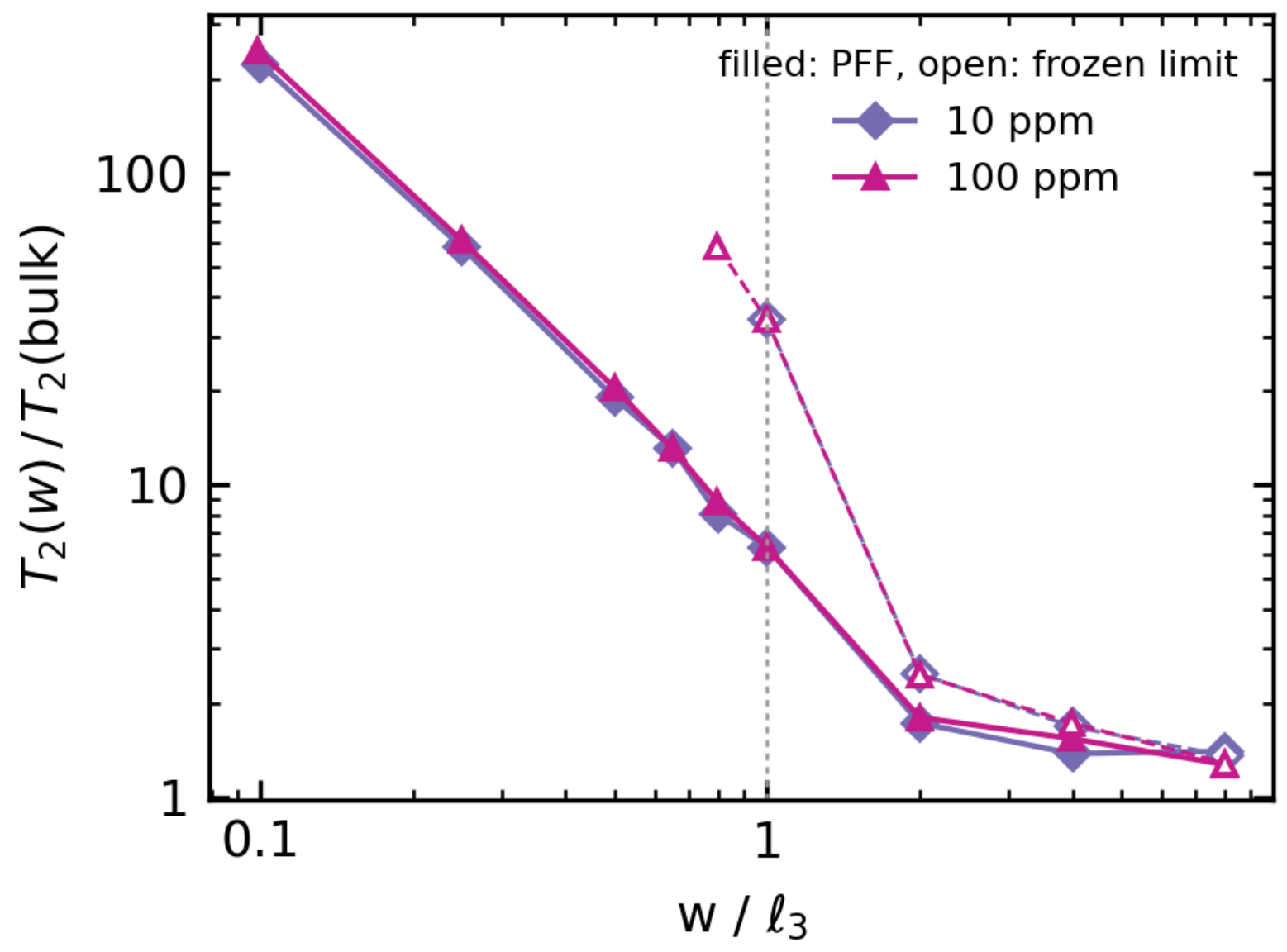

FIG. S5. Echo time relative to bulk of the same local concentration, $T_2(w)/T_2(\mathrm{bulk})$, versus $w/\ell_3$ for 10 ppm (diamonds) and 100 ppm (triangles); PFF (filled) and frozen limit (open, where $|L| = 1/e$ is reached in the window). Dotted: $w = \ell_3$. Bulk $T_2$ is 12.4 µs at 10 ppm and 1.21 µs at 100 ppm.

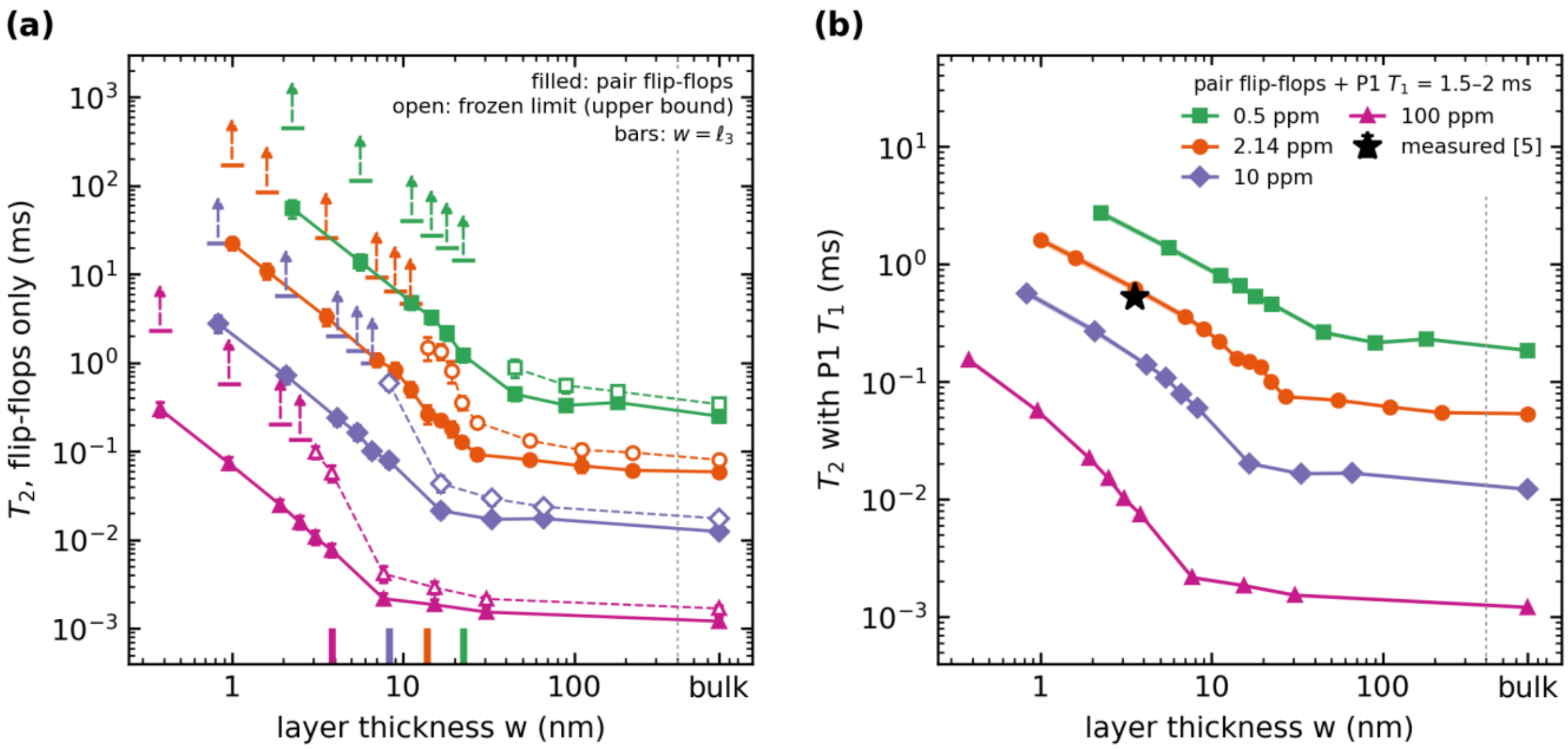

FIG. S6. Main-text Fig. 2 with the 100 ppm series (triangles) added. Echo time ($|L| = 1/e$) versus layer thickness for local concentrations of 0.5, 2.14, 10 and 100 ppm; bulk at the right. Bars at the bottom mark $w = \ell_3$. (a) P1 flip-flops only: PFF (filled) and frozen limit (open). Arrows: lower bounds where the echo stays above 1/e to the end of the computed window. Error bars: bootstrap over configurations. (b) PFF including spin–lattice relaxation of the P1 spins with $T_1$ = 1.5–2 ms (bands) [16]. Star: measured $T_2$ of Ref. [5].

## S6. Hyperfine control

Supplementary Fig. S7 compares P1 centres with bare electrons (no hyperfine, $\langle S_z \rangle = \pm 1/2$), for the sheet and the 110 nm layer at σ = 7.6 ppm·nm. Without hyperfine, the PFF echo time of the sheet falls from 3.2 to 0.52 ms, and that of the 110 nm layer from 2.6 to 0.77 ms; in the frozen limit the sheet echo time falls from beyond the window ($|L| = 0.6$ at $50\tau_0$) to 3.9 ms. The slow 2D bath therefore requires both reduced dimensionality and the static disorder supplied by the hyperfine sub-groups.

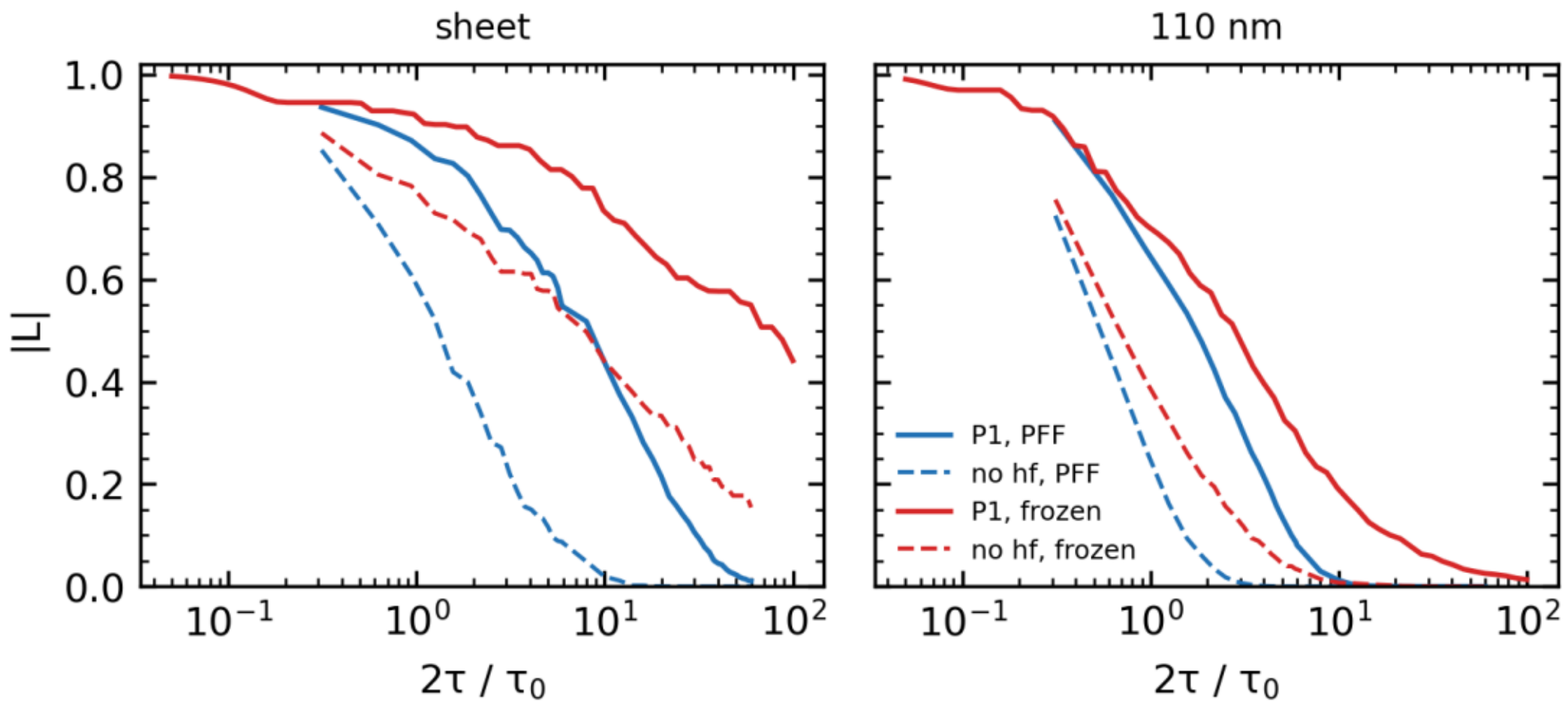

FIG. S7. Echo with P1 centres (solid) and with bare electrons without hyperfine (dashed), PFF (blue) and frozen limit (red), for the sheet and the 110 nm layer, σ = 7.6 ppm·nm.

**Thickness series without hyperfine.** The bare-electron calculation was repeated for the 2.14 ppm series of main-text Fig. 2, at the positions of the P1 series (eight configurations; PFF, short windows of $0$–$3\tau_0$ and $0$–$10\tau_0$). The values $T_2 c_\ell$ (ms·ppm) are:

| $w/\ell_3$ | PFF, no hyperfine | no mean field, no hyperfine | PFF, P1 |
|---|---|---|---|
| 0.07 | 0.146 | 0.064 | 0.92 |
| 0.12 | 0.129 | 0.064 | 0.92 |
| 0.26 | 0.150 | 0.066 | 0.92 |
| 0.51 | 0.143 | 0.064 | 0.83 |
| 0.65 | 0.134 | 0.062 | 0.93 |
| 0.80 | 0.148 | 0.071 | 0.77 |
| 1.01 | 0.121 | 0.062 | 0.57 |
| 1.20 | 0.137 | 0.075 | 0.48 |
| 1.40 | 0.094 | 0.061 | 0.39 |
| 1.59 | 0.078 | 0.046 | 0.275 |
| 1.95 | 0.051 | 0.036 | 0.197 |
| 3.97 | 0.053 | 0.034 | 0.172 |
| 7.94 | 0.047 | 0.032 | 0.147 |
| bulk | 0.051 | 0.030 | 0.126 |

Bootstrap errors are about 20–40 % for thin layers without hyperfine (PFF) and about 10 % in bulk. The mean over the five layers with $w = 0.07–0.65\ell_3$ relative to bulk, with configurations resampled jointly, is 2.7 ± 0.8 (PFF) and 2.1 ± 0.3 (no mean field) without hyperfine, and 7.2 ± 1.4 and 2.9 ± 0.4 with P1. The ratio of the PFF step to the step without mean field is 1.3 ± 0.3 without hyperfine and 2.5 ± 0.4 with P1. Removing the mean field thus leaves a step of 2–3, which we attribute to geometry; a further slowing is resolved only for P1.

We define the midpoint of the crossover as the thickness at which $T_2c_\ell$ equals the geometric mean of the thin-layer and bulk values, interpolated linearly in log w and log $T_2c_\ell$. It lies at $w = 1.47 \pm 0.11\ell_3$ for P1 (16–84 % range of the bootstrap 1.41–1.54) and at $1.50\ell_3$ (1.39–1.64) without hyperfine; the difference, $-0.03\ell_3$ (−0.15 to 0.06), is not significant. Without the mean field the midpoints are 1.64 (1.57–1.79) and 1.66 (1.56–1.76)$\ell_3$, so the position of the crossover changes little with either ingredient. Among the series at fixed local concentration, only the 2.14 ppm series has layers between $\ell_3$ and $2\ell_3$. The fixed-dose series (7.6 ppm·nm) has one, at $1.59\ell_3$ (local concentration 0.14 ppm), with $T_2c_\ell = 0.261 \pm 0.024$ ms·ppm, against 0.275 ± 0.031 at the same $w/\ell_3$ for 2.14 ppm.

## S7. Resonance counting

P1 centres (322 G) are assigned to four spectral sub-groups by the high-field hyperfine shift $m_I A_{zz}$ ($A_{zz}$ = 160.0 MHz for a Jahn–Teller axis along the field, 120.0 MHz otherwise; $m_I = \pm 1/2$). All electron spins are given random $S_z = \pm 1/2$, and each spin feels the static field $h_i$ of all others. A pair (i, j) is resonant if it belongs to one sub-group and satisfies Eq. (5) of the main text, the mutual contribution being removed from both fields. For a P1 at the centre of a sheet (radius 40ℓ, lengths in units of $\ell_2$) or of a bulk sphere (radius 14ℓ), we count resonant partners within R, averaged over 200 configurations. In 3D the count grows as ln R (0.38, 0.50, 0.59 and 0.66 within 2, 4, 6 and 10ℓ); in 2D it saturates at 0.25 by 3ℓ. Without sub-groups the counts are 0.90 (2D, saturated) and 2.2 (3D, at 10ℓ, still growing).

Equation (6) follows from the probability that a pair at distance r is resonant, which for $|J| \ll W$ is about $|J(r)|/W$. Summing over partners with density $n_d$ gives $\int n_d\,(J_0/Wr^3)\,d^d r$, which diverges logarithmically for d = 3 and converges for d = 2. In a layer of thickness w, partners beyond $r \approx w$ are distributed in two dimensions, so the logarithm is cut off at $r \approx w$; the crossover is complete when the cut-off reaches the nearest-neighbour distance, $w \approx \ell_3$.

## S8. Longitudinal relaxation of the P1 spins

Each P1 electron is an independent two-state Markov process, $S_z = \pm 1/2$, with rate $1/(2T_1)$ in each direction, so that $\langle S_z \rangle$ relaxes as $\exp(-t/T_1)$. The echo factor in Eq. (7) is evaluated exactly from equilibrium populations (dynacce.relaxation) and agrees with the analytic 2 × 2 transfer-matrix result to $10^{-15}$. The factors of all P1 centres are multiplied with the PFF echo, configuration by configuration and with the same positions. Relaxation and flip-flops are treated as independent; the change of mean field caused by a jump is neglected.

For bulk diamond, $T_1$ = 1.5–2 ms lowers the PFF echo time from 250 to 179–190 μs at 0.5 ppm, from 59 to 53–54 μs at 2.14 ppm, and from 12.4 to 12.1–12.2 μs at 10 ppm. For the 3.55 nm, 7.6 ppm·nm layer the configuration-averaged echo time is 277, 486, 583, 666, 804 and 1337 μs for $T_1$ = 0.3, 1, 1.5, 2, 3 and 10 ms; between 0.3 and 3 ms, $T_2 \propto T_1^{0.46}$.

**P1 $T_1$.** Reynhardt et al. [16] measured P1 spin–lattice relaxation by pulsed EPR at 9.6 GHz from 4.2 to 300 K. In two synthetic type-Ib crystals containing only P1 centres (25 and 95 ppm), the room-temperature rate is about 500 and 650 $s^{-1}$ ($T_1 \approx$ 2 and 1.5 ms); their fits $1/T_1 = AT + BT^5$ (A = 1.2–5.6 × $10^{-3}$ $s^{-1}K^{-1}$, B = 1.4–5.9 × $10^{-10}$ $s^{-1}K^{-5}$ for the central and hyperfine lines) give 0.7–2.9 ms at 300 K. Above 80 K the $BT^5$ term, attributed to spin–orbit phonon-induced tunnelling of the Jahn–Teller distortion, dominates. Main-text Fig. 4(b) combines these fits with the computed $T_2(T_1)$; for the central line of the 25 ppm crystal it gives 0.67, 1.0 and 1.6 ms at 300, 250 and 200 K.

## S9. Experimental data read from Ref. [5]

The experimental points in main-text Fig. 4(a) were not taken from a data file; they were read from the plot of the single-NV echo in Fig. 3(c) of Hughes et al. [5] (labelled $T_2$ = 540 μs) and replotted on our own axes. Marker centroids were located by colour and converted with the axis frame (2τ = 0–1000 μs, coherence −0.1 to 1.2). Of the plotted points, 120 well-separated markers were recovered; an overlay confirms their positions. The values are provided as a data file; the original data are those of Ref. [5].

## S10. Comparison with Ref. [7]

Schätzle et al. used a partitioned CCE with a static mean-field average, layers perpendicular to [111] and higher field. For a 110–120 nm layer at about 70 ppb their $T_2$ is about 4.5 ms. Our 110 nm layer at σ = 7.6 ppm·nm (0.07 ppm) gives 4.7 ms in the frozen limit, which corresponds to their calculation, and 2.6 ms with PFF.

## Supplementary tables

TABLE S1. Echo time (|L| = 1/e) versus thickness at fixed local concentration (main-text Fig. 2; 100 ppm: Fig. S6). PFF and frozen limit: P1 flip-flops only; ">" gives the end of the computed window when |L| > 1/e there. With $T_1$: PFF including P1 spin–lattice relaxation with $T_1$ = 1.5 and 2 ms. Errors: bootstrap over configurations.

| local [N] (ppm) | w (nm) | w/ℓ₃ | PFF (μs) | frozen limit (μs) | PFF, $T_1$ = 1.5 ms (μs) | PFF, $T_1$ = 2 ms (μs) |
|---|---|---|---|---|---|---|
| 0.5 | 2.25 | 0.1 | 55,695 ± 12,578 | > 452,404 | 2,543 ± 715 | 2,897 ± 732 |
| 0.5 | 5.62 | 0.25 | 13,835 ± 2,664 | > 114,603 | 1,290 ± 199 | 1,461 ± 205 |
| 0.5 | 11.24 | 0.5 | 4,806 ± 803 | > 40,518 | 751 ± 66 | 847 ± 71 |
| 0.5 | 14.62 | 0.65 | 3,260 ± 566 | > 27,314 | 618 ± 45 | 687 ± 49 |
| 0.5 | 17.99 | 0.8 | 2,157 ± 388 | > 20,010 | 499 ± 43 | 562 ± 51 |
| 0.5 | 22.49 | 1 | 1,220 ± 209 | > 14,320 | 433 ± 36 | 481 ± 41 |
| 0.5 | 44.97 | 2 | 444 ± 76 | 874 ± 193 | 250 ± 27 | 275 ± 31 |
| 0.5 | 89.94 | 4 | 331 ± 33 | 555 ± 99 | 207 ± 20 | 223 ± 21 |
| 0.5 | 179.88 | 8 | 356 ± 42 | 473 ± 38 | 222 ± 19 | 238 ± 19 |
| 0.5 | bulk | ∞ | 250 ± 20 | 340 ± 26 | 179 ± 12 | 190 ± 14 |
| 2.14 | 1 | 0.072 | 22,261 ± 3,522 | > 172,439 | 1,496 ± 291 | 1,717 ± 277 |
| 2.14 | 1.6 | 0.12 | 10,895 ± 2,123 | > 85,203 | 1,046 ± 152 | 1,226 ± 181 |
| 2.14 | 3.55 | 0.26 | 3,295 ± 703 | > 25,781 | 580 ± 61 | 662 ± 72 |
| 2.14 | 7 | 0.51 | 1,082 ± 179 | > 9,311 | 332 ± 31 | 383 ± 31 |
| 2.14 | 8.97 | 0.65 | 829 ± 162 | > 6,419 | 266 ± 26 | 298 ± 28 |
| 2.14 | 11.04 | 0.8 | 503 ± 94 | > 4,701 | 210 ± 18 | 232 ± 23 |
| 2.14 | 14 | 1 | 265 ± 63 | 1,480 ± 434 | 153 ± 22 | 164 ± 27 |
| 2.14 | 16.56 | 1.2 | 226 ± 32 | 1,349 ± 276 | 144 ± 12 | 153 ± 15 |
| 2.14 | 19.32 | 1.4 | 181 ± 37 | 811 ± 221 | 130 ± 8 | 137 ± 10 |
| 2.14 | 22.08 | 1.6 | 128 ± 15 | 355 ± 65 | 97 ± 7 | 102 ± 8 |
| 2.14 | 27 | 1.9 | 92 ± 14 | 213 ± 21 | 73 ± 8 | 76 ± 9 |
| 2.14 | 55 | 4 | 80 ± 8 | 132 ± 16 | 68 ± 6 | 71 ± 6 |
| 2.14 | 110 | 7.9 | 69 ± 11 | 104 ± 18 | 60 ± 9 | 62 ± 9 |
| 2.14 | 220 | 16 | 61 ± 7 | 97 ± 6 | 54 ± 5 | 55 ± 5 |
| 2.14 | bulk | ∞ | 59 ± 5 | 80 ± 6 | 53 ± 4 | 54 ± 5 |
| 10 | 0.83 | 0.1 | 2,782 ± 583 | > 22,576 | 531 ± 58 | 604 ± 66 |
| 10 | 2.07 | 0.25 | 719 ± 146 | > 5,732 | 253 ± 31 | 288 ± 36 |
| 10 | 4.14 | 0.5 | 236 ± 40 | > 2,027 | 134 ± 14 | 147 ± 16 |
| 10 | 5.38 | 0.65 | 164 ± 31 | > 1,368 | 104 ± 14 | 112 ± 15 |
| 10 | 6.62 | 0.8 | 100 ± 16 | > 1,002 | 76 ± 9 | 81 ± 10 |
| 10 | 8.28 | 1 | 78 ± 12 | 593 ± 100 | 58 ± 9 | 62 ± 9 |
| 10 | 16.57 | 2 | 21.4 ± 3.4 | 43 ± 9 | 20.0 ± 3.0 | 20.3 ± 3.1 |
| 10 | 33.14 | 4 | 17.1 ± 1.7 | 29.3 ± 5.0 | 16.4 ± 1.8 | 16.6 ± 1.6 |
| 10 | 66.27 | 8 | 17.4 ± 2.0 | 23.7 ± 2.0 | 16.6 ± 2.0 | 16.8 ± 2.0 |
| 10 | bulk | ∞ | 12.4 ± 1.1 | 17.5 ± 1.4 | 12.1 ± 1.0 | 12.2 ± 1.0 |
| 100 | 0.38 | 0.099 | 299 ± 58 | > 2,305 | 145 ± 18 | 161 ± 23 |
| 100 | 0.96 | 0.25 | 74 ± 12 | > 574 | 55 ± 7 | 58 ± 7 |
| 100 | 1.92 | 0.5 | 24.8 ± 3.6 | > 203 | 22.1 ± 3.0 | 22.7 ± 3.2 |
| 100 | 2.5 | 0.65 | 15.8 ± 2.7 | > 137 | 14.9 ± 2.3 | 15.1 ± 2.5 |
| 100 | 3.07 | 0.8 | 10.7 ± 1.9 | 98 ± 16 | 10.1 ± 1.7 | 10.2 ± 1.7 |
| 100 | 3.84 | 1 | 7.63 ± 1.30 | 57 ± 12 | 7.40 ± 1.29 | 7.45 ± 1.31 |
| 100 | 7.69 | 2 | 2.17 ± 0.33 | 4.15 ± 0.87 | 2.16 ± 0.33 | 2.16 ± 0.33 |
| 100 | 15.37 | 4 | 1.85 ± 0.26 | 2.90 ± 0.47 | 1.85 ± 0.24 | 1.85 ± 0.25 |
| 100 | 30.74 | 8 | 1.53 ± 0.17 | 2.16 ± 0.15 | 1.53 ± 0.15 | 1.53 ± 0.16 |
| 100 | bulk | ∞ | 1.21 ± 0.11 | 1.68 ± 0.12 | 1.21 ± 0.10 | 1.21 ± 0.11 |